\documentclass[aps,prl,twocolumn,showpacs]{revtex4}

\usepackage{latexsym}
\usepackage[pdftex]{graphicx}

\begin{document}

\title{ Mechanism of phase transitions and the electronic density of states in (La,Sm)FeAsO$_{1-x}$F$_x$ from {\it ab initio} calculations}

\author{Peter V. Sushko,$^{1,2,*}$ Alexander L. Shluger,$^{2}$ Masahiro Hirano,$^{3}$ Hideo Hosono$^{3}$}
\email[Corresponding author. Email address: ]{p.sushko@ucl.ac.uk}
\affiliation{$^1$WPI-Advanced Institute for Materials Research, Tohoku University, Sendai, Japan}
\affiliation{$^2$Department of Physics {\&} Astronomy, London Centre for 
Nanotechnology, Materials Simulation Laboratory, University College London, 
Gower Street, London WC1E 6BT, United Kingdom}
\affiliation{$^3$Frontier Collaborative Research Center and Materials and Structures Laboratory, Tokyo Institute of Technology, 4259 Nagasuta, Midori-ku, Yokohama 226-8503, Japan}

\begin{abstract}

The structure and electronic density of states in layered LnFeAsO$_{1-x}$F$_x$ (Ln=La,Sm; $x$=0.0, 0.125, 0.25) are investigated using density functional theory. For the $x$=0.0 system we predict a complex potential energy surface, formed by  close-lying single-well and double-well potentials, which gives rise to
the tetragonal-to-orthorhombic structural transition, appearance of the magnetic order, and an anomaly in the specific heat capacity observed experimentally at temperatures below $\sim$140--160~K. We propose a mechanism for these transitions and suggest that these phenomena are generic to all compounds containing FeAs layers. For $x>$0.0 we demonstrate that transition temperatures to the superconducting state and their dependence on $x$ correlate well with the calculated magnitude of the electronic density of states at the Fermi energy. 

\end{abstract}


\date{\today}

\maketitle

The discovery of a new superconductor LaFeAsO$_{1-x}$F$_{x}$ with a high transition temperature ($T_c$=26~K)~[\onlinecite {2008_Kamihara_JACS_26K}] has triggered a global search for other Fe-based alternatives to Cu-based superconductors, which have dominated the field since their discovery in 1986~[\onlinecite{Muller_highTc_1986}]. Substituting As, Fe, and La  for other 
pnicogens~[\onlinecite{Hosono_2007_LaFeOP}], 
transition metals~[\onlinecite{Hosono_2007_LaNiOP}] and 
lanthanides~[\onlinecite{SmFeAsO_Nature_43K}], respectively, 
applying external pressure~[\onlinecite{Hosono_2008_Nature_pressure}], and optimizing the doping level have pushed the $T_c$ to 54.5~K~[\onlinecite{GdFeAsO_2008_55K}]. However, since then its value seems to have saturated. As doubts have been expressed that $T_c$ can be raised any further [\onlinecite{Grant_2008_Nature}], it became apparent that generic guiding principles for the $T_c$ optimization need to be developed. 

\begin{figure}[htbp]
\begin{center}
\includegraphics[angle=0,width=8.5cm]{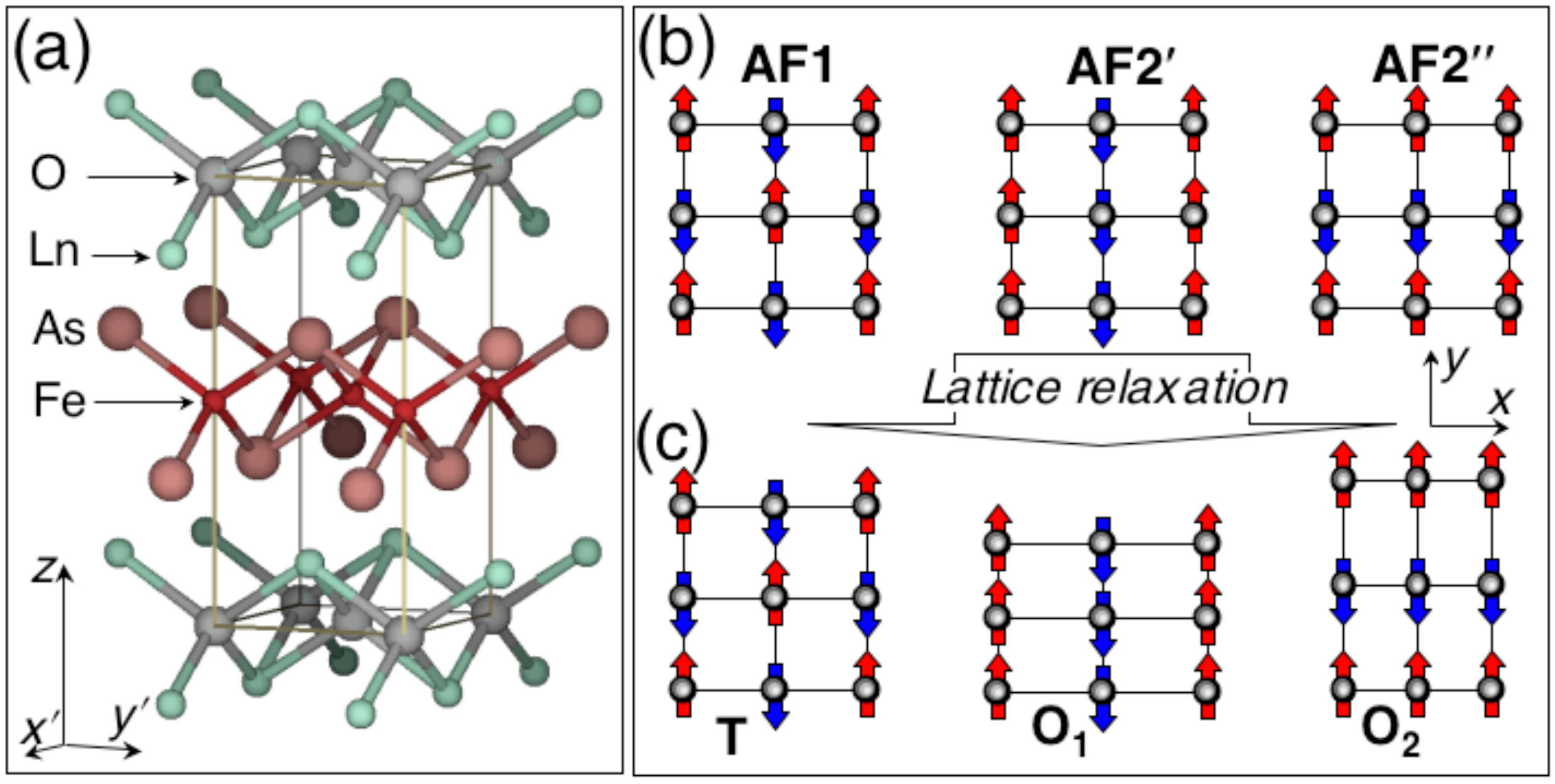}
\caption{(Color online) (a) Structure of 1$\times$1 LnFeAsO (Ln = La, Sm) unit cell. 
(b) Schematics of several spin configurations within Fe layers shown for a $\sqrt{2}\times\sqrt{2}$ supercell. The circles show the positions of Fe atoms within the Fe layer.  Here and below, up and down arrows indicate "up" and "down" spins, respectively. See text for details.}
\label{fig_struct}
\end{center}
\end{figure}

LaFeAsO is a member of the layered Fe-pnicogens, in which FeAs sheets are separated from each other by spacers such as layers of ionic oxide, e.g. LaO in LaFeAsO, [Fig.~\ref{fig_struct}(a)] or metal atoms, 
e.g. Ba in BaFe$_2$As$_2$ [\onlinecite{Rotter_BaFe2As2}].  
In spite of the difference in the nature of the spacers, FeAs-based materials show surprising similarities in the temperature dependence of their structural parameters, anomalies in the electric resistance and specific heat capacity, and in their magnetic properties (e.g.~[\onlinecite{Nomura_structure,Kohama_phase_trans,McGuire_phase,Margadonna_SmFeAsO,Rotter_BaFe2As2}]).

A series of theoretical and computational reports appeared recently  describing the electronic properties, magnetic interactions, phonon structure, and the origin of the superconductivity in LaFeAsO and related compounds (e.g.~[\onlinecite{Haule_2008_PRL_LaFeAsOF,Du_2008_PRL_LaFeAsO,Terakura_2008_JPSJ_LaFeAsO,Mazin_LaFeAsO,Aoki_LaFeAsO,Yildirim_LaFeAsO}]). The aim of this work is twofold: (i) to develop a model for the  phase transitions observed in FeAs-based materials and (ii) to investigate a correlation between electronic density of states at the Fermi energy and the experimentally observed values of the $T_c$ and its dependence on the doping level. 

The calculations were carried out using density functional theory (DFT), the generalized gradient approximation functional PW91~[\onlinecite{Perdew_1992_PRB}] and the projected augmented waves method~[\onlinecite{Blochl_1994_PRB_PAW}] implemented in the VASP code~[\onlinecite{Kresse_1996_PRB_VASP}]. 
The plane-wave basis set cutoff was set to 600 eV. The supercells containing eight (1$\times$1, Fig.~\ref{fig_struct}), 16 ($\sqrt{2}$$\times$$\sqrt{2}$), and 32 (2$\times$2) atoms and Monkhorts-Pack grids of 252, 132, and 36 $k$-points, respectively, were used. For the analysis of the electronic structure, the charge-density was decomposed over atom-centered spherical harmonics. 

In the first part of the paper we consider the relation between configurations of the spins associated with Fe 3$d$ electrons and the lattice structure. Several ordered antiferromagnetic configurations in a $\sqrt{2}$$\times$$\sqrt{2}$ supercell are shown in Fig.~\ref{fig_struct}(b). In AF1, the spins on the neighboring Fe atoms are antiparallel. In configurations AF2$'$ and AF2$''$ spins are parallel along $y$- and $x$-axes respectively; AF2$'$ and AF2$''$ are equivalent in the case of the high-temperature tetragonal ({\bf T}) phase.

\begin{figure}[htbp]
\begin{center}
\includegraphics[angle=0,width=8.5cm]{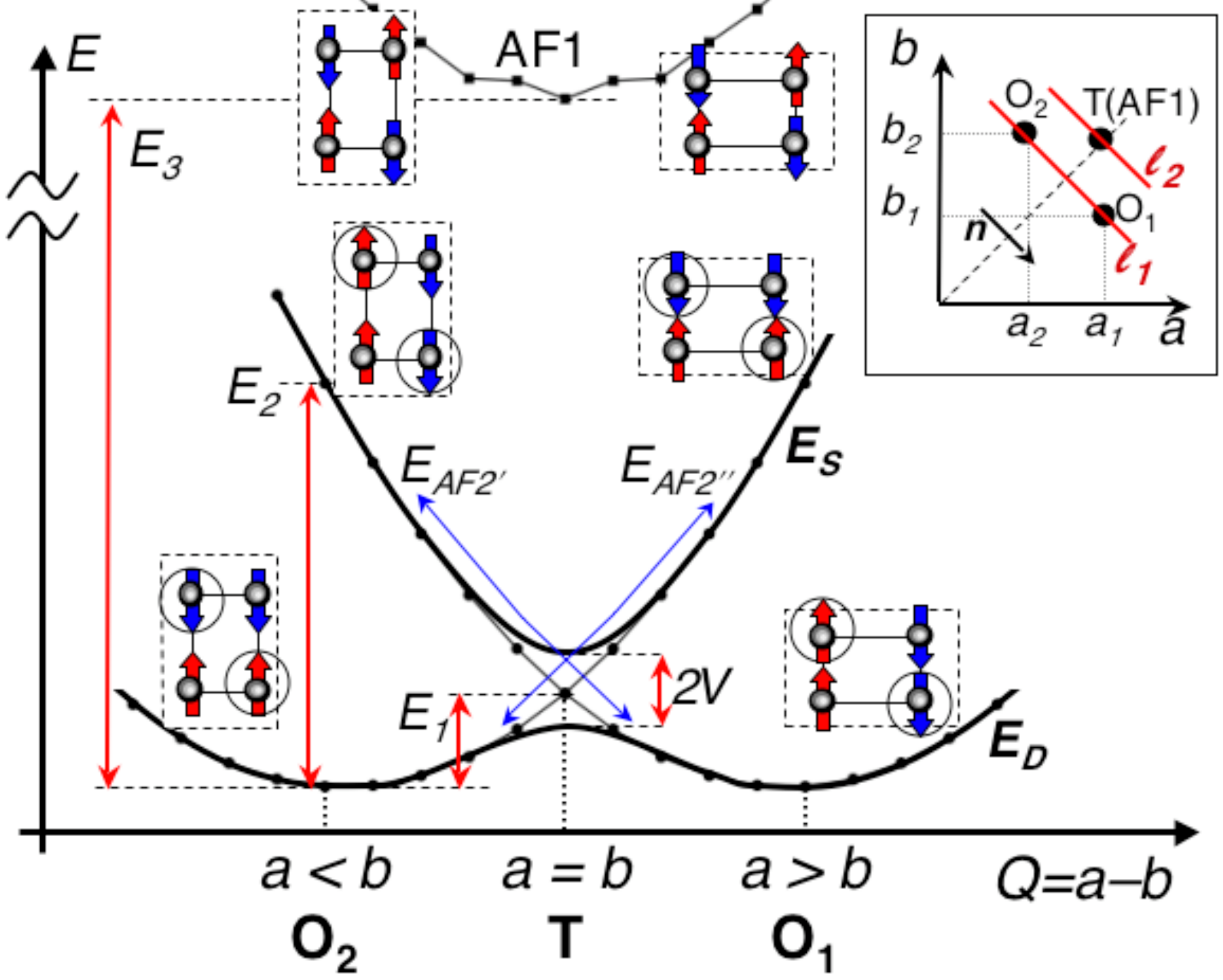}
\caption{Potential energy surfaces for the AF1 and AF2 configurations. Dots correspond to calculated energy values. Open circles indicate the spin pairs, which are different in {\bf O$_1$} and {\bf O$_2$} configurations.}
\label{fig_eng}
\end{center}
\end{figure}

After minimization of the total energies with respect to both the atomic positions and the lattice parameters, the AF1 configuration maintains the {\bf T} structure (Table~\ref{tbl_struct}). Configurations AF2$'$ and AF2$''$ relax to two equivalent orthorhombic ({\bf O}) structures {\bf O$_1$} and {\bf O$_2$}, in which the Fe 3$d$ spins along the short Fe--Fe bonds are parallel and those along the long Fe--Fe bond are antiparallel [see Fig.~\ref{fig_struct}(c)]. The lattice parameters $a$, $b$, and $c$ for {\bf O$_1$} and {\bf O$_2$} relate as $a_1$=$b_2$, $b_1$=$a_2$, $c_1$=$c_2$, and $a_1$$>$$b_1$ (see also Fig.~\ref{fig_eng}). The calculated values for the lattice parameters for the low-temperature {\bf O}-phase agree with the experimental data to within 0.4~\%. (Table~\ref{tbl_struct}). The ferromagnetic configuration is considerably less stable than antiferromagnetic ones and is not considered here. 

Integration of  the AF2$''$ charge-density within LaO and FeAs layers shows that the layers are charged: (LaO)$^{+\delta}$(FeAs)$^{-\delta}$ with $\delta$ = 0.15 $|e|$. Thus, one can consider LnFeAsO as a super-ionic compound, in which ionic and ion-covalent bonding within the LnO and FeAs layers, respectively, is accompanied by the weak ionic bonding of these layers. The magnetic moments on Fe atoms calculated for AF2 are 1.56~$\mu_{B}$ (Ln=La) and 1.33~$\mu_{B}$ (Ln=Sm). These differ significantly from the values suggested by M\"ossbauer measurements ($\sim$0.35~$\mu_B$)~[\onlinecite{Kitao_Mossbauer}]. 

To find the energy barrier separating the fully relaxed AF2$'$ and AF2$''$ configurations, we calculated the total energies $E_{AF2'}$ and $E_{AF2''}$ along the path $\ell_1$ connecting {\bf O$_1$} and {\bf O$_2$}  (inset in Fig.~\ref{fig_eng}). Path $\ell_1$ is parallel to the vector $\mathbf{n}$=(1,--1) in the $a$-$b$ plane. The $E_{AF2'}(Q)$ and $E_{AF2''}(Q)$, where $Q$=$a$--$b$, are plotted in Fig.~\ref{fig_eng}. For comparison, we also calculated $E_{AF1}(Q)$  along the path $\ell_2 || \mathbf{n}$. 

The calculated values of $E_1$, $E_2$, and $E_3$ are 0.005 eV, 0.025 eV, and 0.15 eV, respectively, for LaFeAsO and 0.006 eV, 0.026 eV and 0.11 eV for SmFeAsO. We note that approximate exchange-correlation functionals, such as PW91, can underestimate the values of 
energetic characteristics by as much as 100\%. More reliable values of $E_1$, $E_2$, and $E_3$, as well as those of Fe magnetic moments, can be obtained by applying methods, which include exact exchange interaction and allow for coupling of different many-electron 
states~[\onlinecite{Hozoi_2007_CAS_periodic}]. 

At $Q$=0, AF2$'$ and AF2$''$ have the same atomic structures and $E_{AF2'}$=$E_{AF2''}$, yet, their electronic states are different. This situation leads to Jahn-Teller (JT) instability~[\onlinecite{book_Stoneham_defects}] and formation of a conical intersection at the crossover of the potential energy surfaces (PESs) $E_{AF2'}$ and $E_{AF2''}$.
Correcting for non-adiabatic behavior near the intersection, together with taking into account the coupling of many-electron states, introduces an effective interaction $V$, which splits the $E_{AF2'}$ and $E_{AF2''}$ into a higher-energy single-well potential ($E_S$) and a lower-energy double-well potential ($E_D$)~[\onlinecite{book_Stoneham_defects}] 
as shown in Fig.~\ref{fig_eng}. We can conservatively estimate that $0<V<E_1$. 

The lattice dynamics, described by $E_S$ and $E_D$, has three regimes depending on the temperature ($T$): 

1. For $T < E_1 - V$, the atoms vibrate near the their positions defined by one of the orthorhombic energy minima of $E_D$ (e.g.~{\bf O$_1$}). In this case the magnetic structure is dominated by AF2$'$ configuration (see Fig.~\ref{fig_eng}).

2. For $E_1 - V < T < E_1 + V$, motion of atoms is determined by parabolic branches of $E_D$, although the effect of the barrier separating its energy minima can not be neglected. The difference between the average distribution of short and long Fe--Fe bonds decreases with increasing temperature, which corresponds to a gradual transition from {\bf O} to {\bf T} symmetry. Magnetic order is lost because the Fe spins adjust themselves to the momentary local atomic structure, so as the spins are parallel for Fe atoms forming short Fe--Fe bonds and anti-parallel otherwise. In other words, thermal fluctuations of Fe--Fe bond lengths cause reorientation of Fe spins (Fig.~\ref{fig_eng}).

3. For $T>E_1+V$, the lattice dynamics is determined by parabolic branches of $E_S$ and $E_D$ and the effect of the barrier in $E_D$ can be neglected. The lattice has the {\bf T}-symmetry. There is no magnetic order because the orientation of the spins changes according to the local atomic structure, as described in 2, and also due to coupling of electronic states of $E_S$ and $E_D$.

Experimental observations of the structural and magnetic phase transitions in LaFeAsO (e.g. [\onlinecite{Cruz_2008_structure,Nomura_structure,McGuire_phase}]) suggest that the {\bf T}$\rightarrow${\bf O} transition takes place gradually, with the $Q$=$a$--$b$ order parameter exhibiting two kinks at $T_{max}$ ($\sim$160~K) and $T_{min}$ ($\sim$140~K), and that the magnetic phase transition occurs at $T_{min}$ or slightly below it. In addition, specific heat capacity displays two peaks, which also seem to coincide with $T_{max}$ and $T_{min}$~[\onlinecite{Kohama_phase_trans,McGuire_phase}]. Similar data have been reported for other FeAs-based materials~[\onlinecite{Rotter_BaFe2As2}]. These results are consistent with the model for the three regimes of the lattice dynamics outlined above, in which two phase transition temperatures $T_{max}$ and $T_{min}$ correspond to $E_1 + V$ and $E_1 - V$, respectively. 
We can also speculate, that the decrease in the amplitude of atomic vibrations during {\bf T}$\rightarrow${\bf O} transition~[\onlinecite{book_Zaiman}] can contribute to the abrupt drop in the electrical resistivity observed, for example, in [\onlinecite{2008_Kamihara_JACS_26K}].

\begin{figure}[htbp]
\begin{center}
\includegraphics[angle=0,width=8.5cm]{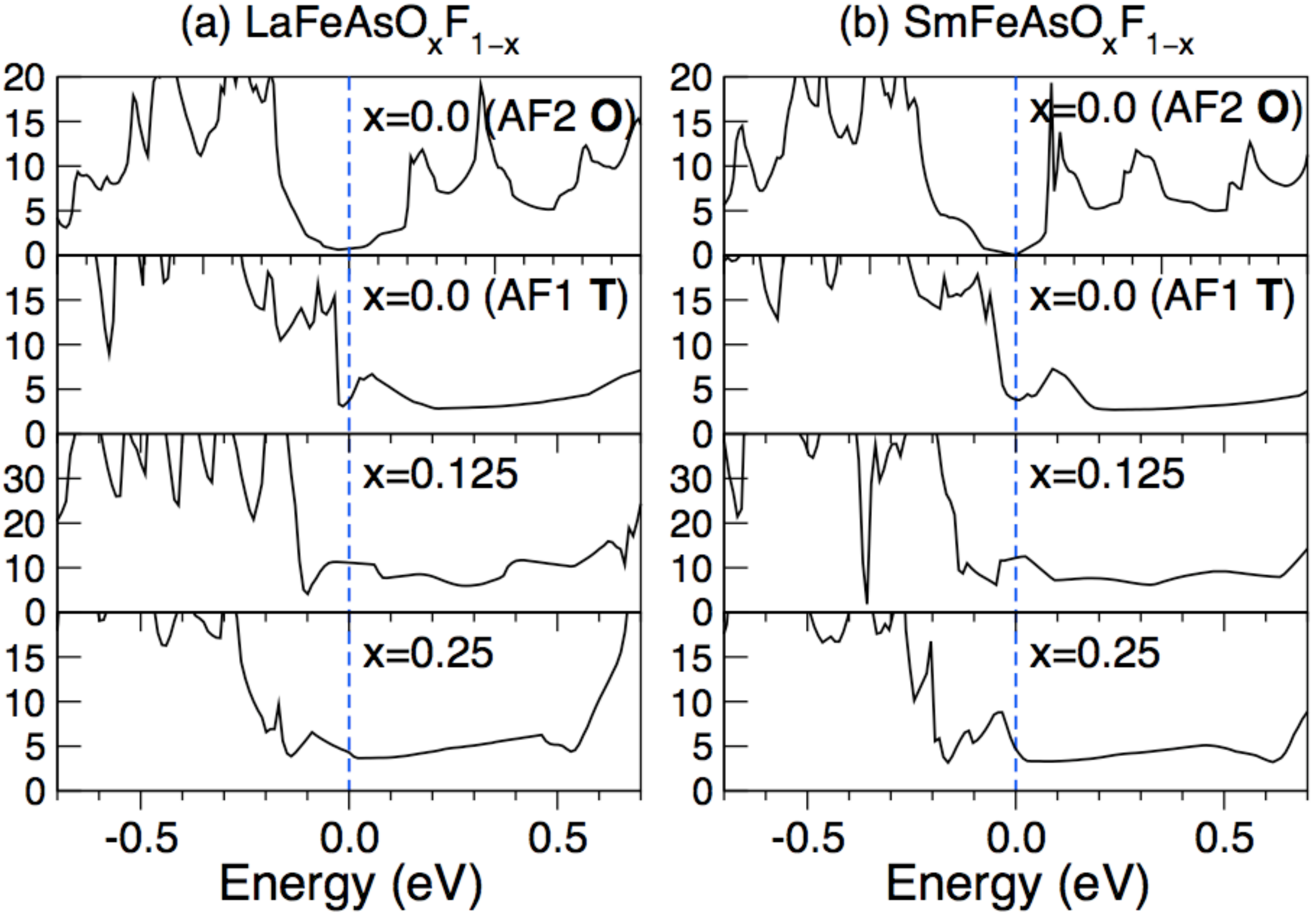}
\caption{Density of states for (La,Sm)FeAsO$_{x}$F$_{1-x}$. Letters {\bf O} and {\bf T} refer to the orthorhombic and tetragonal phases, respectively. The Fermi energy is at 0.0~eV.}
\label{fig_doped}
\end{center}
\end{figure}

We now consider the effect of F-doping on the atomic and electronic structures of LnFeAsO. The doping provides additional electrons to the FeAs layer so as the charge distribution becomes (LnO)$^{+\delta+x}$(FeAs)$^{-\delta-x}$ and the lattice parameter $c$ decreases due to the increased inter-layer ionic bonding (Table~\ref{tbl_struct}). This leads to opening up of a narrow gap in the $N(\varepsilon)$ at $\sim$2.5 eV below the $\varepsilon_F$ (not shown). 

We find that the spin-density distribution in the FeAs layers is not independent on the arrangement of the F impurities. For example, for $x$=0.25 ($\sqrt{2}\times\sqrt{2}$ cell), the spin-down density is localized on a single Fe atom nearest to the F$^-$ impurity, while the remaining three Fe atoms share the spin-up density. 
At the doping level of $x$=0.125 (2$\times$2 cell), the effect is more subtle. The lowest energy state is similar to that of the undoped LnFeAsO: the lattice structure corresponds to {\bf O}-symmetry of the $\sqrt{2}\times\sqrt{2}$ cell and the spin-arrangement is the same as in AF2, although the values of $\mu_{Fe}$ are reduced to 1.32 (Ln=La) and 0.75 $\mu_{B}$ (Ln=Sm). We also found a spin-disordered state, which has the {\bf T}-symmetry and is $\sim$7 (Ln=La) and $\sim$5 (Ln=Sm) meV per Fe atom higher than the ground state. Taking into account the generally random distribution of the F impurities over O lattice sites in realistic samples, we suggest that such spin-disordered state realizes in practice.

For $x$=0.125 (2$\times$2 cell) we distinguish two sets of non-equivalent Ln and As atoms with different values of their $z$-coordinates. The effect of such structure on the lattice phonons and on the charge- and spin-density distributions needs to be considered separately.

Finally, we investigate the correlation between the structure and doping level and the electronic density of states [$N(\varepsilon)$] calculated for the fully relaxed AF1, AF2, and doped LnFeAsO$_{1-x}$F$_{x}$ (Fig.~\ref{fig_doped}). In all cases the $N(\varepsilon)$ near the Fermi energy ($\varepsilon_F$) is dominated by the Fe 3$d$ states and the polarization of spin-up and spin-down states is negligible. 

In stoichiometric LnFeAsO, $N_{AF2}(\varepsilon)$ has a pronounced depression near $\varepsilon_F$, while the $N_{AF1}(\varepsilon)$ has a narrow deep minimum separating a steep rise at $\varepsilon<\varepsilon_F$ and a peak at $\varepsilon$$>$$\varepsilon_F$~[\onlinecite{comment_AF2}]. Projecting $N_{AF1}(\varepsilon)$ on the $d$-states shows that this peak is dominated by $d_{xz}$ and $d_{yz}$ states. The same $d_{xz}$+$d_{yz}$ peaks near $\varepsilon_F$ are evident for the doped LnFeAsO~(Fig.~\ref{fig_doped}).

\begin{table}[htdp]
\caption{Structural parameters of LnFeAsO$_{1-x}$F$_x$ (Ln=La,Sm). In all cases crystallographic cell angles $\alpha$, $\beta$, and $\gamma$ deviate from 90$^{\circ}$ by less than 0.0005$^{\circ}$. Letters $E$ and $T$ refer to experiment and theory (this work) respectively.}
\begin{center}
\begin{tabular}{l|cc|ccc|cc}
\hline
 $x$ & \multicolumn{2}{c|}{details} & $a$,~\AA & $b$,~\AA & $c$,~\AA & $z$(Ln) & $z$(As) \\
 \hline
\multicolumn{8}{l}{LaFeAsO$_{1-x}$F$_{x}$}  \\
 0.0    & AF1    & $T$ & 5.6873 & 5.6899 & 8.6185 & 0.1448 & 0.6383 \\
 0.0    & AF2    & $T$  & 5.7305 & 5.6672 & 8.6948 & 0.1433 & 0.6438 \\
 0.0    & 300 K & $E$~[\onlinecite{Nomura_structure}]  &
                                5.7031 & 5.7031 & 8.74111 &  0.1413 & 0.6517 \\
 0.0    & 120 K & $E$~[\onlinecite{Nomura_structure}]  &
                                5.6826 & 5.7104 & 8.71964 &  0.1417 & 0.6513 \\
0.125 &            & $T$ & 5.6829 & 5.6829 & 8.5630 & 0.1560 & 0.6405 \\
           &           &  &             &             &             & 0.1452 & 0.6394 \\
 0.25   &           & $T$ & 5.6873 & 5.6831 & 8.4859 & 0.1562 & 0.6410 \\
 0.14   &120 K  & $E$~[\onlinecite{Nomura_structure}] &
                                5.6844& 5.6844 & 8.6653 & 0.1477 & 0.6527 \\
\hline
\multicolumn{8}{l}{SmFeAsO$_{1-x}$F$_{x}$}  \\
 0.0     &   AF1   & $T$ & 5.5955 & 5.5918 & 8.3435 & 0.1406 & 0.6472 \\
 0.0     &   AF2   & $T$ & 5.6232 & 5.5623 & 8.4142 & 0.1396 & 0.6515 \\
 0.125 &             & $T$ & 5.5834 & 5.5834 & 8.2884 & 0.1523 & 0.6496 \\
           &             &  &             &             &             &  0.1413 & 0.6479 \\
 0.25   &             & $T$ & 5.5888 & 5.5902 & 8.2046 & 0.1529 & 0.6493 \\
\hline
\end{tabular}
\end{center}
\label{tbl_struct}
\end{table}

According to the standard BCS theory of superconductivity, the transition temperature $T_c$ is proportional to $\left< \omega \right> exp[-1/\lambda N(\varepsilon_F)]$, where $\left< \omega \right> $ is a typical phonon frequency and $\lambda$ is the electron-phonon coupling constant. As shown in Ref. [\onlinecite{Abrikosov_2008}], the limitation of $T_c <$40~K, suggested by Migdal's theorem for BCS superconductors, is not justified and, therefore, much higher values of the $T_c$ can be achieved by optimizing $\left< \omega \right>$, $\lambda$, and $N(\varepsilon_F)$. We can tentatively suggest that $\left< \omega \right>$ and $\lambda$ do not vary strongly for FeAs-based compounds, since the conductivity is confined to the FeAs layers. Then $T_c$ can be considered as a function of a single parameter $N(\varepsilon_F)$.  

Thus, we consider the correlation between the behavior of $N(\varepsilon)$ for $\varepsilon$ close to $\varepsilon_F$ (Fig.~\ref{fig_doped}) and experimentally observed properties of  LnFeAsO$_{1-x}$F$_{x}$ superconductors. 
First, we notice that as $x$ increases and $\varepsilon_F$ shifts across the $d_{xz}$+$d_{yz}$ peak, the value of $N(\varepsilon_F)$ increases as well, then reaches its maximum and then decreases. The details of the peak structure depend of the value of $x$ but its general shape is reminiscent of the experimentally observed dependence of the $T_c$ on $x$ (e.g.~[\onlinecite{2008_Kamihara_JACS_26K,Margadonna_SmFeAsO}]).

Furthermore, the maximum of the $d_{xz}$+$d_{yz}$ peak ($x$=0.0) in SmFeAsO is higher and further away from $\varepsilon_F$ than that in LaFeAsO.  This correlates with the  observations that the optimal $T_c$ is higher in SmFeAsO$_{1-x}$F$_{x}$ (46~K, $x$=0.15~[\onlinecite{SmFeAsO_Nature_43K}]) than in LaFeAsO$_{1-x}$F$_{x}$ (26~K, $x$=0.05--0.12~[\onlinecite{2008_Kamihara_JACS_26K}]) and that it is achieved at larger values of $x$. The slope of $N(\varepsilon_F)$ calculated for $x$=0.125 is negative for Ln=La and positive for Ln=Sm, which indicates that maximum of $N(\varepsilon_F)$ can be found at $x<$0.125 for Ln=La and $x>0.125$ for Ln=Sm.  This is consistent with the optimal values of $x$ found for these compounds as $\sim$0.11 (Ln=La)~[\onlinecite{2008_Kamihara_JACS_26K}] and $\sim$0.20 (Ln=Sm)~[\onlinecite{Margadonna_SmFeAsO}].

Finally, we notice that the $d_{xz}$+$d_{yz}$ peak in LaFeAsO is wider than that in SmFeAsO (this is clearly seen for $x$=0.0 and 0.125), which suggests that $T_c$ has a stronger dependence on $x$ in SmFeAsO as observed in~[\onlinecite{Margadonna_SmFeAsO}]. While these observations say little about the mechanism of the superconductivity in FeAs-based materials, they suggest that the highest $T_c$ can be achieved in those, which have the largest magnitude of the $d_{xz}$+$d_{yz}$ peak close to $\varepsilon_F$.

To summarize, we investigated the PESs for different magnetic states of stoichiometric LnFeAsO (Ln=La,Sm) and found that the properties of this system are determined by two close-lying PESs: a lower-energy double-well potential, where each well corresponds to the orthorhombic symmetry, and a higher-energy single-well potential of the tetragonal symmetry. This complex potential energy surface gives rise to three temperature ranges, and, therefore, two transition temperatures, and can explain the experimentally observed structural phase transition, the appearance of the magnetic order, and the anomaly in the temperature dependence of the specific heat capacity. 

We noticed a correlation between the calculated profile of $N(\varepsilon)$ near $\varepsilon_F$ and experimentally observed dependence of the $T_c$ on the dopant concentration $x$ and on the type of Ln atom. This correlation can be used for computational prescreening of the promising LnFeAsO derivatives as well as for predicting optimal dopant concentrations via relatively inexpensive electronic structure calculations.

The authors thank C. R\"uegg and A. M. Stoneham for their comments on the manuscript and S. W. Kim, Y. Kamihara, T. Nomura, and T. Kamyia for valuable discussions. P. V. S. is grateful to Japan Science Foundation and  WPI-AIMR at Tohoku University. The access to HPCx is provided via the Materials Chemistry Consortium.

\end{document}